\documentclass[12pt]{article}
\usepackage{amsmath,amsthm,amscd,amsfonts,amssymb,mathrsfs}
\usepackage{latexsym}
\usepackage{graphics}
\usepackage{color}
\usepackage[german,english]{babel}  
\usepackage[latin1]{inputenc} 
\usepackage{graphicx, color}
\usepackage{amssymb}
\newtheorem{theorem}{Theorem}
\newtheorem{proposition}[theorem]{Proposition}
\newtheorem{lemma}[theorem]{Lemma}
\newtheorem{corollary}[theorem]{Corollary}
\newtheorem{definition}{Definition}

\newtheorem{remark}[theorem]{Remark}

\newtheorem{hyp}{Hypotheses}
\newcommand{\R}{\mathbb{R}}

\newcommand{\Z}{\mathbb{Z}}

\renewcommand{\P}{\mathbb{P}}
\newcommand{\E}{\mathbb{E}}

\newcommand{\F}{\mathcal{F}}
\renewcommand{\H}{\mathcal{H}}

\newcommand{\Rd}{\mathbb{R}^{d}}

\def\half{{1 \over 2}}
\def\<{\langle}
\def\>{\rangle}

\def\p0{\psi_0}                              

\def\1Ll{{1 \over {\vert \Lambda_L \vert}}}

\def\j1d{{(1+\vert j \vert )}^{-(d+2)}}
\def\x1a{{(1+ \vert x\vert )}^{-\alpha}}
\def\2j1a{{(1+ \vert j\vert )}^{-\alpha}}
\newcommand{\es}{\quad\par\vspace{0.25cm}}
\newcommand{\tn}[1]{\textnormal{#1}}

\newcommand{\PP}{\mathbb P}
\newcommand{\RR}{\mathbb R}

\newcommand{\ZZ}{\mathbb Z}
\newcommand{\EE}{\mathbb E}

\newcommand{\hh}{\mathcal H}
\newcommand{\ii}{\mathcal I}
\newcommand{\Ii}{\mathscr I}

\newtheorem{thm}{Theorem}[section]

\def\pf{ {\noindent \bf Proof: } }
\begin{document}
\date{31 March 2009}
\author{W Kirsch \\  Facult\"at f\"ur Mathematik und Informatik\\  Fern Universit\"at in Hagen \\
58084 Hagen, Germany \\ and \\ M Krishna \\Institute of Mathematical Sciences
\\ Taramani, Chennai 600113, India }
\title{Lifshitz tails for the Interband Light Absorption Coefficient}
\maketitle
\begin{abstract}
In this paper we consider the Interband Light Absorption Coefficient 
for various models.  We show that at the lower and upper edges of the spectrum
the Lifshitz tails behaviour of the density of states implies similar behaviour
for the ILAC at appropriate energies.  The Lifshitz tails property is also exhibited at some
points corresponding to the internal band edges of the density of states.
\end{abstract}

\section{Introduction}

In this work we look for Lifshitz tails behaviour of the Interband Light Absorption Coefficient (ILAC) defined in equation (eqnl4). The standard definition
of the ILAC involves considering a pair of operators of the form
$H_\omega^\pm = \Delta \pm V^\omega$, with $\Delta$ the Laplacian on either $\ell^2(\Z^d)$, in the
discrete case or on $L^2(\Rd)$ in the continuous case, and taking a random
potential $V^\omega$.  Restricting these operators $H_\omega^\pm$ to 
boxes $\Lambda$ gives operators with discrete spectra so that in any finite region of
energy these operators have only finitely many eigenvalues.  Using this fact
one can define the quantity
$$
\frac{1}{Vol(\Lambda)} \sum_{\lambda_\omega^- + \lambda_\omega^+ \leq E} |\langle \phi_{\omega,\lambda_\omega^-}, \psi_{\omega,\lambda_\omega^+} \rangle |^2
$$
where $\phi_{\omega,\lambda_\omega^-}, \psi_{\omega,\lambda_\omega^+}$ are the eigen functions of the operators $H_\omega^\mp$ restricted to the box $\Lambda$, corresponding to the eigenvalues 
$\lambda_\omega^-, \lambda_\omega^+$ respectively.

The limit of the above quantity, when it exits, gives the ILAC.  

We consider a correlation measure (mentioned also in \cite{KhoKirPas})
$\rho$ and identify the ILAC as the distribution function of a marginal
of the measure $\rho$ in a diagonal direction.  This identification enables
us to prove theorems on the Lifshitz tails behaviour of the ILAC more easily
since it involves only comparing the marginal of $\rho$ with the density
of states of either of the  operators $H_\omega^\pm$.  We also do not
need to approximate to define the ILAC, but can obtain the function directly.

In the next section, we present an abstract version of the correlation measure
$\rho$ and the density of states $n$ for a pair of random covariant operators 
and obtain relations between the two.

\section{General Covariant Operators}

We start with a definition of  a random family of self adjoint operators
which are covariant under a group action.

\begin{hyp}\label{hyp1}\es
\begin{enumerate}
\item $\H$ is a (separable, complex) Hilbert space, $(\Omega,\F,\P)$ a probability space.
\item There is a locally compact abelian group $G$ and $\{U_x\}_{x\in G}$ is a group of unitary operators on $\H$, 
i.e. the $U_x$ are unitary and $U_{x+y}=U_x\,U_y$, $U_0=\textnormal{Id}$, $U_{-x}=U_x^{-1}=U_x^*$
\item There is a discrete subgroup $L$ of $G$ and an orthogonal projection $P$ on $\H$ such that $\{U_n^*PU_n\}_{n\in L}$, $\{U_nPU_n^*\}_{n\in L}$ are orthogonal partitions of unity on $\H$. We set $P_n=U_n^* P U_n, \tilde{P}_n = U_n PU_n^*.$
\item $\{T_n\}_{n\in L}$ is a group of probability preserving transformations on $\Omega$.
\end{enumerate}
\end{hyp}

\begin{definition}
A family $\{A_\omega\}_{\omega\in\Omega}$ of self-adjoint operators on $\H$ is called 
measurable if the family $\{(A_\omega+i)^{-1}\}_{\omega\in\Omega}$ is measurable
\end{definition}

It is known (see \cite{CFKS}, \cite{CarLac} and section 2.4 of \cite{IvaVes}) that a family of \emph{bounded} self-adjoint operators is measurable iff it's
weakly measurable.

Moreover, if $\{A_\omega\}$ is a measurable family of self-adjoint operators then for any bounded measurable
function $f$ the operator family $f(A_\omega)$ is weakly measurable. (also in \cite{CFKS}, \cite{CarLac}, section 2.4 \cite{IvaVes}).

Finally, the product of weakly measurable families is weakly measurable (see \cite{CarLac}).

\begin{definition}
A weakly measurable family $A_\omega$ of bounded operators is called \emph{covariant} (with respect to
$U_x,T_x$) if
$$A_{T_x\omega}=U_x^*\,A_\omega\,U_x\qquad \tn{for all $x\in G$} $$

Also, a measurable family $A_\omega$ of self adjoint operators is called \emph{covariant} (with respect to
$U_x,T_x$) if
$$A_{T_x\omega}=U_x^*\,A_\omega\,U_x\qquad \tn{for all $x\in G$} $$
\end{definition} 

If $A_\omega$ is a covariant family of self-adjoint operators and $f$ is a bounded measurable function, then
the family $f(A_\omega)$ is covariant (also in \cite{CFKS}, \cite{CarLac}). Moreover, if both $A_\omega$ and $B_\omega$ are covariant
families of bounded operators, then $A_\omega\,B_\omega$ is a covariant family. We denote by
$\|B\|_1$ the trace norm of a trace class operator $B$.

\begin{proposition}\label{propl2}
Let $A_\omega$ and $B_\omega$ be covariant families of bounded operators and assume
that $A_\omega P$ and $B_\omega P$ are trace class and 
\begin{equation}\label{eqnl5}
\E(\|A_\omega P\|_1)<\infty ~ \mathrm{and} ~ \E(\|B_\omega P\|_1)<\infty.
\end{equation}

Then:
\begin{equation}\E(Tr (PA_\omega B_\omega P))~=~\E(Tr(P B_\omega A_\omega P))\label{equ:comm}\end{equation}
\end{proposition}

\begin{proof}
\begin{align}
Tr(P A_\omega B_\omega P) ~&=~Tr(P A_\omega B_\omega P)\\
&=~ \sum_n\; Tr(P A_\omega  P_n B_\omega P)\\
\intertext{since $P_n$ is a partition of unity of orthogonal projections.}
&=~ \sum_n\; Tr(  P_n B_\omega P A_\omega P_n)\\
\intertext{using the $Tr(AB) = Tr(BA)$ and the invariance of trace $Tr(U^*CU) = Tr(C)$,}
&=~ \sum_n\; Tr(P A_{T_n^{-1}\omega} \tilde{P}_n B_{T_n^{-1}\omega} P)\\
\intertext{ from the covariance of $A_\omega$ and $B_\omega$ }
&=~  Tr( P B_{T_n^{-1}\omega} A_{T_n^{-1}\omega}P) \rangle
\end{align}
In the last step we used the fact that $\tilde{P}_n$ is a partition of unity also.
Now we take expectations of either side of the above equation and obtain
\begin{align}
\E(Tr(P A_\omega B_\omega P)~&=~\E(\sum_n\; Tr(P B_{T_n^{-1}\omega} \tilde{P}_n A_{T_n^{-1}\omega}P))\\
&=~\sum_n\; \E(Tr(P B_{T_n^{-1}\omega}\tilde{P}_n A_{T_n^{-1}\omega}P))\\
\intertext{We have used Fubini's theorem to interchange expectation and sum, allowed because of (\ref{eqnl2})}
&=~\sum_n\; \E(Tr(P B_{\omega} \tilde{P}_n A_{\omega}P))\\
\intertext{since $T_n^{-1}$ is probability preserving}
&=~\E(\sum_n\; Tr(P B_{\omega}\tilde{P}_n A_{\omega}P))\\
\intertext{$\tilde{P}_n$ is a partition of unity.}
&=~\E(Tr( B_\omega A_\omega P))~=~\E(Tr(PB_\omega A_\omega P))
\end{align}
\end{proof}

\begin{corollary}\label{corl1}
\begin{enumerate}
\item If $A_\omega, B_\omega, C_\omega$ are covariant families of bounded operators satisfying the condition (\ref{eqnl5}) then:
\begin{equation}
\E(Tr(PA_\omega B_\omega C_\omega P))~=~\E(Tr(PC_\omega A_\omega B_\omega P))
\end{equation}
\item If $A_\omega, B_\omega$ are covariant families of bounded, positive (i.e. $\geq 0$) operators satisfying the conditions (\ref{eqnl5}) then
\begin{equation}
\E(Tr P A_\omega B_\omega P))~\geq~0
\end{equation}
\end{enumerate}
\end{corollary}
\begin{proof}
The first assertion is clear as we can apply the proposition to the covariant families $A_\omega B_\omega$ and $C_\omega$.

For the second claim we observe that $B_\omega=C_\omega C_\omega$ with a $C_\omega=\sqrt{B_\omega}$. $C_\omega$ as a function of
the covariant family $B_\omega$ is covariant as well. Moreover, since $A_\omega$ is positive (and the Hilbert space is complex),
$A_\omega$ is self-adjoint and so is $C_\omega$.

By part (i) of the corollary we have:
\begin{align}
\E(Tr(P A_\omega B_\omega P) )~&=~\E(P A_\omega C_\omega C_\omega P))\\
&=~\E(Tr(P C_\omega A_\omega C_\omega P))\\
&\geq~0 \qquad\tn{since $A_\omega$ is positive}
\end{align}

\end{proof}

\begin{hyp}\label{hypl2}
Let $H_\omega$ be family of self-adjoint operators, which are bounded below, on a Hilbert space $\hh$.  
Let $ E_{_{H_\omega} } (\cdot)$ be the (projection-valued) spectral measure of $H_\omega$ such that 
for any bounded borel set $A$, the operators $P E_{_{H_\omega} }(A), E_{_{H_\omega} }(A) P$ 
are trace class for a.e. $\omega$ and form a {\bf covariant} family of operators.
\end{hyp}

For operators $H_\omega$ satisfying the above hypothesis, it is clear that
for any finite $x$, the spectral measure $ E_{_{H_\omega} }( (-\infty, x ] ) =
 E_{_{H_\omega} }( [c , x ] )$, with $c$ finite and smaller than the
 infimum of the spectrum of $H_\omega$.  Therefore the hypothesis implies
 that for any finite $x$, the operators $P E_{_{H_\omega} }( (-\infty, x ] ),
 E_{_{H_\omega} }( (-\infty, x ] ) P$
 are trace class. Therefore we can now define the density of states
 for such operators.

\begin{definition}\label{defl0}
 Let $H_\omega$ be a family of self adjoint operators satisfying 
 Hypothesis \ref{hypl2}.  Then the \emph{density of states} of this 
 family is defined to be the unique $\sigma$-finite measure $n$
 associated with the monotone right continuous function $F$, 
$$
F(x) = \E\left( Tr(P E_{_{H_\omega} }( (-\infty, x ] ) P)  \right),
$$
via $n( (a, b] ) = F(b) - F(a), ~ a, b \in \R. $ 
\end{definition}
Thus for any bounded borel set $A$, $n(A)$ agrees with the right hand side
of the above relation with $A$ replacing $(-\infty, x]$.

In the above framework we define another measure that is used to define
the Interband Light Absorption Coefficient (ILAC).  To do this we need a
 pair $H_\omega^\pm$ of self-adjoint operators as in 
the Hypothesis \ref{hypl2} and consider the associated projection
valued measures $E_{H_\omega^\pm}(\cdot)$.  We then define the density
of states of these operators by,  
\begin{equation}\label{eqnl00}
n_\pm(A) = \EE \left( Tr ( P E_{H_\omega^\pm}(A) P ) \right).
\end{equation}

Consider the semi algebra $\ii \times \ii$ of subsets of $\R^2$ where
$$
\ii = \RR \cup \{(a, b]: a, b \in \RR\} \cup \{(a, \infty) : a \in \RR\} \cup  \{(-\infty, a] : a \in \RR\}.$$

We define the correlation measure $\rho$ on $\ii \times \ii$ as
\begin{equation}\label{eqnl3}
\rho(A\times B) = \EE \left( Tr(P E_{H_\omega^+}(A) E_{H_\omega^-}(B)P  ) \right),
\end{equation}
where $\rho$ is set to be $\infty$ if either $A$ or $B$ is an unbounded element of $\ii$.

This set function takes values in $[0,1]$ if $P$ is trace class and in $[0, \infty]$, if $PE_{H_\omega^\pm( (a,b] )}$ are trace class only for bounded intervals $(a,b]$, in view of Proposition \ref{propl1}. We set
$$
\rho(A) = \sum_{i=1}^\infty \rho(A_i\times B_i), ~ \mathrm{if} ~ A = \sqcup_{i=1}^\infty A_i, A_i \in \ii. 
$$
It is a simple exercise to see that this $\rho$ is well defined on $\ii \times \ii$ and via standard
measure theory extends as a $\sigma$-finite measure to the whole borel $\sigma$-algebra of $\R^2$..

Using the Hypothesis 2, and Proposition \ref{propl2} we see that the
following is valid. 

\begin{proposition}\label{propl1}
Consider the operators $H_\omega^\pm$ satisfying Hypothesis \ref{hypl2}
and let $n_\pm$ and $\rho$ be as in equation (\ref{eqnl3}). 
Then for any $B, C \in \ii$ bounded,
\begin{enumerate}
\item 
$\rho(B\times C) = \EE \left( Tr( P  E_{H_\omega^-}(C) E_{H_\omega^+}(B) E_{H_\omega^-}(C) P) \right)$
\item 
$\rho(B\times C) = \EE \left( Tr( P E_{H_\omega^+}(B) E_{H_\omega^-}(C) E_{H_\omega^+}(B) P ) \right)$
\item The following inequalities are valid
$$
\rho(B\times C) \leq n_+(B), ~ \rho(B \times C) \leq n_-(C).
$$
\end{enumerate} 
\end{proposition}
\pf Since the subsets $B, C$ are bounded the operators $PE_{H_\omega^-}(C),  PE_{H_\omega^+}(B)$ are covariant trace class operators satisfying the inequality
(\ref{eqnl5}).  Therefore the result follows by an application of Proposition
\ref{propl2} and Corollary \ref{corl1}.  \qed

We collect the arguments about $\rho$ in a proposition.

\begin{proposition}\label{propl3}
Consider a pair of covariant operators $H_\omega^\pm$ satisfying the Hypothesis \ref{hypl2} and consider the correlation measures $\rho$ extended to the borel $\sigma$-algebra on $\R^2$ from that
given by equation (\ref{eqnl3}).  Then the following are valid.
\begin{enumerate}
\item If $P$ is trace class, then $\rho$ is a probability measure on $\R^2$,
with support in the closure of $\cup_{\omega} \sigma(H_\omega^+) \times \sigma(H_\omega^-)$.
\item If $P$ is not trace class but, $PE_{H_\omega^\pm}( (a, b] )P$ is trace class,
for bounded intervals $(a, b]$, then $\rho$ is a positive $\sigma$-finite 
measure on $\R^2$, with support in the closure of $\cup_{\omega} \sigma(H_\omega^+) \times \sigma(H_\omega^-)$.
\end{enumerate}
\end{proposition}

\begin{remark}\label{reml1}
Typically the first case occurs for operators on $\ell^2(\Z^d)$ and the second
case occurs in $L^2(\Rd)$. 
\end{remark}

We take the transformation $T$ on $\RR^2$ given by
$$
T\left(\begin{matrix}\lambda_1\\ \lambda_2\end{matrix}\right) = \left(\begin{matrix}\frac{\lambda_1 + \lambda_2}{\sqrt{2}}\\ \frac{\lambda_1 - \lambda_2}{\sqrt{2}}\end{matrix}\right).
$$
Using this $T$ we define the Interband Light Absorption Coefficient (ILAC) $A$ 
as the distribution function, 
\begin{equation}
\label{eqnl4}
A(\lambda) - A(\lambda^\prime) = \nu\left( \frac{1}{\sqrt{2}}(\lambda^\prime, \lambda]\right), ~ \mathrm{where} ~
\nu(B) = \rho \circ T^{-1} (B\times \RR)
\end{equation}
In the above equation the factor $\frac{1}{\sqrt{2}}$ is because of the normalisation we used for $T$, so that this definition of ILAC agrees with the standard
one in the case of finite box operators.  We also note that since the operators
$H_\omega^\pm$ are assumed to be bounded below $A(-\infty) = 0$.

In the case when $\PP$ in Hypothesis \ref{hyp1} is ergodic with respect to
the action of $G$ on $\Omega$, then, the spectra $\sigma(H_\omega^\pm)$ of covariant families of
operators $H_\omega^\pm$ are almost everywhere constant sets.  In such
a case we can talk about the infimum of spectra of $H_\omega^\pm$ without
reference to $\omega$.  In this context we have the following theorem.

\begin{thm}\label{thml1}
Suppose $H_\omega^\pm$ are a pair of random families of self-adjoint
operators satisfying Hypothesis \ref{hyp1}. Assume further that
$\PP$ is ergodic with respect to the action of $G$ on $\Omega$.
\begin{enumerate}
\item Let $E_\pm = \inf \sigma(H_\omega^\pm)$.  
Then $A(E_+ + E_- + a) - A(E_+ + E_--a) \leq n_{\pm}( (E_\pm -2a, E_\pm+ 2a) ), 
~ a>0$. 
\item Let $E_\pm^\prime = \sup \sigma(H_\omega^\pm)$.  
Then $A(E_+^\prime + E_-^\prime + a) - A(E_+^\prime + E_-^\prime-a) \leq n_{\pm}( (E_\pm^\prime -2a, E_\pm^\prime+ 2a) ), 
~ a>0$. 
\end{enumerate}
\end{thm}
\pf We shall prove the first case, the other proof is similar (where
one has to use the fact that $\lambda_1 \leq E_+^\prime, \lambda_2 \leq E_-^\prime$ respectively for the other case and work it out). 
Let $E_+, E_-$ to be the infima of the spectra 
$\sigma(H_\omega^+), \sigma(H_\omega^-)$ of $H_\omega^+, H_\omega^-$. 
We consider the closure of the Cartesian product  
$\Sigma = \sigma(H_\omega^+) \times \sigma(H_\omega^-)$
of the spectra of $H_\omega^\pm$, which is the support of the measure $\rho$.
Therefore if we denotes points of $\Sigma$ by $(\lambda_1, \lambda_2)$,
so that $\lambda_1 \geq E_+, \lambda_2 \geq E_-$, then
the possible values of
$\lambda_1 + \lambda_2$ have a lower bound $E_- + E_+$, so 
$\lambda_1 + \lambda_2 \in (E_-+E_+, E_-+E_+ + a)$ implies 
$\lambda_1 \in (E_+ -2a, E_+ + 2 a) ~ \mathrm{and} ~ \lambda_2 \in (E_- -2a, E_- + 2 a)$,  
(see Figure 2).
This immediately implies the
inclusions (the first inclusion is clear and the second one uses the above):

\begin{eqnarray*}
& \{(\lambda_1, \lambda_2) :\lambda_2 \in (E_-, E_-+ (a/2)) ~ and ~  \lambda_1 \in (E_+, E_++ (a/2))\}
\\ & \subset \{(\lambda_1, \lambda_2) : \lambda_1 + \lambda_2 \in (E_-+E_+ -a , E_-+E_++a) \} \\ & \subset 
\{(\lambda_1, \lambda_2) :\lambda_2 \in (E_-, E_-+ 2a) ~ and ~  \lambda_1 \in (E_+, E_++ 2a)\}.
\end{eqnarray*}
This then would lead to the inequalities that 
\begin{eqnarray*}
& A(E_++E_-+a) - A(E_++E_-) \\ &
= \rho \circ T^{-1} ( \frac{1}{\sqrt{2}}(E_+ + E_-, E_++E_-+a] \times \R ) 
\\ & = \rho \left(\{(\lambda_1, \lambda_2) : E_-+E_+ \leq \lambda_1 + \lambda_2 \leq  E_-+E_++a\} \right) \\ & \leq 
\rho \left( (E_-, E_-+2a) \times (E_+, E_+ + 2a) \right) \\ & 
\leq min \{ \rho \left( (E_-, E_-+2a) \times \RR \right), 
\rho \left( (E_+, E_++2a) \times \RR \right) \} \\ &
\leq min \{ n_-\left( (E_- -2a, E_-+ 2a) \right),n_+\left( (E_+ -2a, E_++ 2a) \right) \},
\end{eqnarray*}
where the last inequality comes from Proposition \ref{propl1}(3) and enlarging
the intervals slightly, which only increases the bound since $n_\pm$ are
measures. \qed 

\begin{remark}
If the density of states $n_\pm$ have Lifshitz tails behaviour 
$n_\pm( (E_\pm-a, E_\pm + a) ) \approx e^{-C a^\alpha}$ as $a$ goes to zero,
for an appropriate $\alpha$, at 
$E_\pm$ respectively, then we have 
$$
\limsup_{a >0} \frac{1}{h(2a)}n_\pm\left( (E_--2a, E_-+ 2a) \right) < \infty, 
$$
for $h(a) = e^{-C a^\alpha}$ for some $\alpha$, 
so, using the above inequalities, 
\begin{eqnarray*}
& \displaystyle{\limsup_{a >0}} \frac{1}{h(2a)} (A(E_-+E_++a)-A(E_-+E_+-a)) \\ & \leq \displaystyle{\limsup_{a >0}} \frac{1}{h(2a)} n_+ \left( (E_+-a, E_+ + 2a) \right) < \infty.
\end{eqnarray*}
\end{remark}

In the case when the density of states $n_\pm$ have Lifshitz tails behaviour
at other internal band edges, the same behaviour is valid for ILAC under
some conditions.  Suppose the spectra of $H_\omega^\pm$ consist of 
bands $\cup_{i=1}^N [a_i^\pm,b_i^\pm]$.  Then the product of the spectra is 
$\cup_{i=1,j}^N [a_i^+,b_i^+] \times [a_j^-, b_j^-]$.  Let us denote 
$R_{ij} = [a_i^+,b_i^+] \times [a_j^-, b_j^-]$.  Then, the measure
$\rho$ is supported on the set $\cup_{i=1,j}^N R_{ij}$.  

We index the pairs $(ij)$ by $\beta$ and use $R_\beta$ to denote a rectangle
forming part of $\Sigma$ henceforth.  So we have
$\Sigma = \cup_{\beta} R_\beta$.

The central point in the proof of Theorem \ref{thml1} is that if
$(c,d)$ is a corner of the rectangle formed by the lowest bands of the spectra
of $H_\omega^\pm$, then the strip $\{ (\lambda_1, \lambda_2) : c+d \leq \lambda_1 + \lambda_2 \leq c+d+a\}$ intersected with the support of $\rho$ is a triangle
of side length $\sqrt{2} a$, (see Figure 2 ), hence its $\rho$ measure
is smaller than that of the square with the corner $(c,d)$ and side length
$2a$, as can be seen in the Figure 2.  As we see in Figure 1, there may
be some rectangles in the support of $\rho$, with this property.  Those
rectangles in Figure 1, where this is not true are marked by $X$ and the
solid lines are those lines $\lambda_1 + \lambda_2 =const$ for which this 
feature is valid and the dashed lines are those for which this is not true.

In the definition below the sets $R_\beta \subset \RR^2$ and we denote
the coordinates of $\RR^2$ by $(\lambda_1, \lambda_2)$.

\begin{definition}\label{defl1}
Let the support of $\rho$ be $\Sigma = \cup_{\beta} R_\beta$, with 
$R_\beta = [a_i^+,b_i^+] \times [a_j^-, b_j^-], ~ \beta =(ij)$.  
Then we call a corner
$(c,d)$ of a rectangle $R_\beta$ {\bf good}, if the intersection of the line 
$\lambda_1 + \lambda_2 = c+d $ with $\Sigma$ consists of finitely 
many points and all of them are corners of rectangles forming $\Sigma$.  
Given a corner $(c,d)$ in $\Sigma$ we shall denote by $K_{c,d}$ the
set of corners that lie on the line $\lambda_1 + \lambda_2 = c+d $.
\end{definition}

\begin{thm}\label{thml2}
Let spectra of $H_\omega^\pm$ be as in theorem \ref{thml1} and let $\Sigma$
be the support of the measure $\rho$ given in equation \ref{eqnl3}. Let
$A$, as given in equation (\ref{eqnl4}) be the corresponding ILAC. 
If
$(c,d)$ is a good corner in $\Sigma$. Denote the elements of
$K_{c,d}$ by $\{(c_\gamma, d_\gamma) \}$. Then we have
\begin{eqnarray*}
& A(c+d+a) - A(c+d-a) \\& \leq \sum_{(c_\gamma,d_\gamma) \in K_{c,d} } \mathrm{min} \left\{ n_+( (c_\gamma-2a, c_\gamma + 2a) ), n_-( (d_\gamma-2a, d_\gamma + 2a) ) \right\}. 
\end{eqnarray*}
\end{thm}
\pf Firstly we note that if we take a rectangle, $R_\beta$, then only the
lower-left and the top-right corners are candidates of being \emph{good} corners, since for the other two corners, the line $\lambda_1 + \lambda_2 = const$
that contains the said corner will pass through the rectangle and hence has 
infinitely many points.  We will prove the theorem for a good corner $(c,d)$ which is a lower left corner of a rectangle, the proof for the case
of a top-right good corner is similar. In this case we see immediately that 
if $(c,d)$ is a good corner in $\Sigma$, then
the intersection of the strip 
$S_a\left( (c,d) \right) = \{(\lambda_1, \lambda_2): c+d \leq \lambda_1 + \lambda_2 \leq c+d+a \}$ 
with $\Sigma$ is contained in finitely many rectangles $R_\beta$ forming
$\Sigma$.  Further $S_a\left( (c,d) \right)  \cap R_\beta$ is contained
in a square of side length $2a$ contained in $R_\beta$ and having one
corner common with a corner of $R_\beta$.  Given a good
corner $(c,d)$ and the associated strip $S_a\left( (c,d) \right)$,  
let $(c_\gamma, d_\gamma) \in K_{c,d}$ denote the corner of rectangle $R_\gamma$
that has nonempty intersection with it. (Note that this corner
satisfies $c_\gamma + d_\gamma = c+d$).

Then whenever
$(c,d)$ is a good corner we have the inequality, with $\gamma$ ranging
over a finite set, 
\begin{eqnarray}\label{eqnl1}
 S_a\left( (c,d) \right) \cap \Sigma  \subset \cup_{(c_\gamma, d_\gamma) \in K_{c,d}}  [c_\gamma, c_\gamma + 2a]\times [d_\gamma, d_\gamma + 2a].   
\end{eqnarray}

This inequality implies immediately that:
\begin{eqnarray}\label{eqnl2}
& A(c+d+a) - A(c+d -a) \nonumber \\  
& \leq A(c+d+a) - A(c+d) = \rho(S_a\left( (c,d) \right) \cap \Sigma) \nonumber \\
& \leq  \sum_{(c_\gamma, d_\gamma) \in K_{c,d}} \rho\left([c_\gamma, c_\gamma + 2a)\times [d_\gamma,  d_\gamma + 2a)\right) \nonumber \\
& \leq \sum_{(c_\gamma, d_\gamma) \in K_{c,d}} \mathrm{min} \left\{ n_+\left( (c_\gamma-2a, c_\gamma + 2a) \right), n_-\left( (d_\gamma-2a, d_\gamma + 2a) \right)\right\},
\end{eqnarray}
where in the last inequality we enlarged the sets using the fact that $n_\pm$
are measures. 

This shows that at ILAC has the same continuity property as the density of states at the band edges.  \qed

In the theorem below we identify good corners for a simple case of spectra having two bands.

\begin{thm}\label{thml3}
Consider a pair of self adjoint operators $H_\omega^\pm$ as in 
Theorem \ref{thml1}. Suppose a.e. $\omega$, the spectra of 
$H_\omega^+, H_\omega^-$ are given by $\cup_{i=1}^2 [a_i^+, b_i^+]$ and 
$\cup_{i=1}^2 [a_i^-, b_i^-]$, respectively, where 
$a_i^\pm, b_j^\pm$ are listed in the increasing order.  Then the corners 
$$
\{ (a_1^+, a_1^-), (b_1^+, b_1^-), (a_2^+, a_2^-), (b_2^+,b_2^-)   \}
$$
are \emph{good} whenever $a_i^\pm, b_i^\pm$ satisfy, 
\begin{eqnarray*}
&a_1^++a_1^- < b_1^++b_1^- < max(a_2^++a_1^-, a_1^++a_2^-) \\ &
< max(b_2^++b_1^-, b_1^++b_2^-) < a_2^++a_2^- < b_2^++b_2^-.
\end{eqnarray*}
In the case $a_i^+ = a_i^-, b_i^+=b_i^-, i=1,2$, even the corners 
$$\{ (a_2^+, a_1^-), (a_1^+, a_2^-), (b_2^+, b_1^-), (b_1^+, b_2^-)\}
$$ 
are good.
\end{thm}
\pf This is direct verification to see that the diagonal lines $\lambda_1 + \lambda_2 = const$
passing through the respective corners do not intersect any other rectangle.  
In the latter case when the spectra are the same, we have $a_2^++a_1^- = a_1^++a_2^-$
and $b_2^++b_1^- = b_1^++b_2^-$, hence the stated result. \qed

\begin{remark}
In the symmetric case $a_i^\pm = a_i, b_j^\pm = b_j$, however, the rectangles
$R_\beta, R_\gamma \subset S_\beta$ if $\beta = (ij), \gamma = (ji)$. 
In this case the above assumption still ensures that the lower-left and top-right corners of the rectangles are good.
\end{remark}

\newpage

\begin{figure}[htp]
\centering
\input{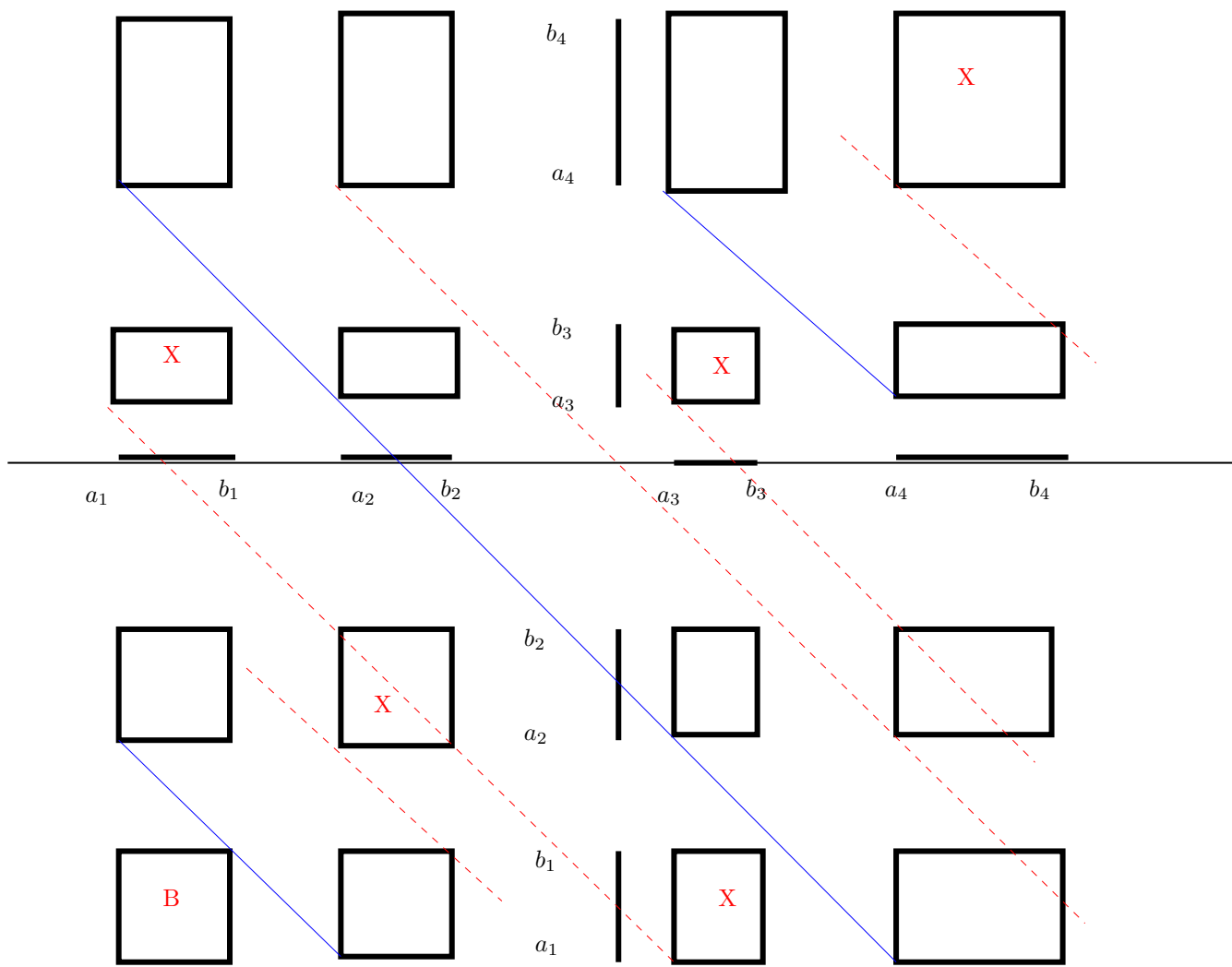}
\caption{Products of Spectra}
\end{figure}

\newpage

\begin{figure}[htp]
\centering
\input{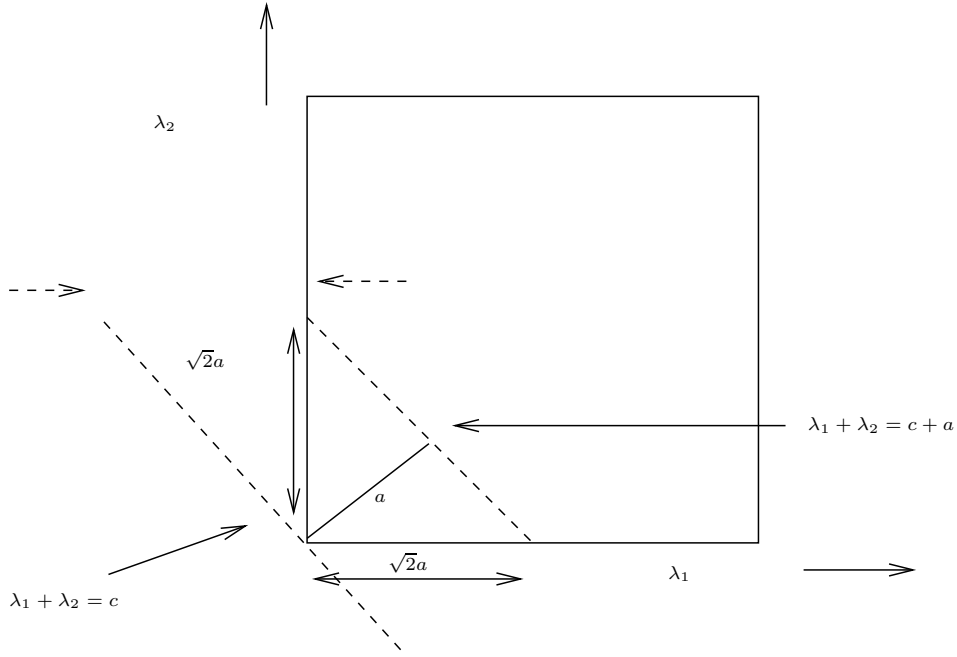}
\caption{A corner of a rectangle}
\end{figure}

\section{Discrete Models:}

Consider $\ell^2(\ZZ^d)$ and the discrete Laplacian
$(\Delta u)(n) = \sum_{|n-i|=1} u(i)$.  Consider real valued
i.i.d random variables $\{q(n) \}$ with common distribution $\mu$.
Let $V_\omega$ denote the operator of multiplication by the sequence $q_\omega(n)$. 
Consider the operators 
$$
H_\omega^\pm = \Delta \pm q_\omega. 
$$
Taking $G=L=\Z^d$, it is known that operators $E_{H_\omega}(A)$ are covariant.
The projection $P$ is taken to be the projection 
$|\delta_0\rangle \langle \delta_0|$ onto the subspace
generated by the vector $\delta_0$, which is an element of the standard basis
for $\ell^2(\ZZ^d)$. 

Then the density of states in these models are given by
$$
n_\pm( (a, b) ) = \EE \left( Tr(PE_{H_\omega^\pm}( (a, b) ) \right)
= \EE \left( \langle \delta_0, E_{H_\omega^\pm}( (a, b) ) \delta_0 \rangle \right)
$$
and the correlation measure $\rho$ is given by
$$
\rho( (a,b) \times (c,d) ) = \EE \left( \langle \delta_0, E_{H_\omega^+}( (a, b) ) E_{H_\omega^-}( (c, d) ) \delta_0 \rangle \right)
$$
and is a probability measure as per Proposition \ref{propl3} (1), since
$P$ is trace class in this case. 

In this model the density of states of $H_\omega^\pm$ are shown to have
Lifshitz tails behaviour at the bottom of the spectra \cite{Sim1},
under the condition that $\mu$ satisfies $\mu( (a, a+\epsilon) ) \geq
C \epsilon^N$, where $a$ is the infimum of the support of $\mu$.   

In the case when the support of $\mu$ has two 
closed intervals $[a_1, b_1] \cup [a_2, b_2]$,($a_i, b_i$ arranged in an increasing
order so that $a_{i+1} > a_i$ for all $i$) and
such that $b_i + 2d < a_{i+1} - 2d$, Simon \cite{Sim2} proves 
the Lifshitz tails behaviour at the internal band edges, if
$\mu$ satisfies $\mu( (a_i, a_i + \epsilon) ) \geq C \epsilon^N$
and $\mu( (b_i -\epsilon, b_i) ) \geq C \epsilon^N$ for all $i$ .  
When $[a_1 -2d , b_1+2d]$ and $[a_2-2d, b_2+2d]$ are disjoint, Lifshitz
tails behaviour at the band edges is also shown for the associated
density of states $n$.  That is at any of the band edges one has
$n( (E-\delta, E+\delta) ) = O( e^{- C \delta^{-\frac{d}{2} } } )$
as $\delta \rightarrow 0$.  

An application of Theorems \ref{thml1}, \ref{thml2} shows that the results are true for 
the ILAC $A$, namely
\begin{thm}\label{thml4}
Consider the Anderson models as above on $\ell^2(\Z^d)$.  
If $[a_1^\pm, b_1^\pm] \cup [a_2^\pm, b_2^\pm]$, are $\pm (supp (\mu) )$.
Then, for some $C>0$, 
\begin{itemize}
\item (External band edge case) For $E \in \{ a_1^+ + a_1^-, b_2^++b_2^-\}$, one has 
$$
A(E+\delta) - A(E-\delta) = o(e^{-C\delta^{-\frac{d}{2}}}), ~ \mathrm{as} ~ \delta \rightarrow 0.
$$
\item (Internal band edge case) If the gap between the intervals $[a_1^\pm-2d, b_1^\pm+2d]$ and $[a_2^\pm-2d, b_2^\pm+2d]$ is large enough, then  
$$
A(E+\delta) - A(E-\delta) = o(e^{-C\delta^{-\frac{d}{2}}}), ~ \mathrm{as} ~ \delta \rightarrow 0,
$$
for any $E \in \{b_1^++a_1^-, a_2^+ + a_1^-, b_2^++a_1^-, a_1^++a_2^-, b_1^++b_2^-, a_2^++a_2^-\}$.
\end{itemize}
\end{thm}

\begin{remark}
We gave a simple example of a discrete model, however there are many more,
those with periodic backgrounds \cite{KirSim}, those which are unbounded
\cite{Klo4} and so on.  We refer to the review \cite{KirMet} for the numerous
cases where the Lifshitz tails for the density of states is proved and for which our theorem applies to the ILAC. 
\end{remark}

\section{Continuous Models:}

Let us start by stating a theorem which is essentially a very weak version
of the uncertainty principle.  

We take $H_0 = -\Delta = -\sum_{i=1}^d D_j^2,
D_j = i\frac{\partial }{ \partial x_j}$, is self adjoint 
on its natural domain
in $L^2(\Rd)$ and its spectrum is $[0,\infty)$. 

We start with a couple of lemmas. We recall the definition of the trace ideal
${\mathscr I}_p$ to be those bounded operators $A$ with the property
$|A|^p$ is trace class. Recall that elements of $\Ii_1$ are called Trace class operators.

\begin{lemma}\label{leml1}
Consider $L^2(\Rd)$ and the operator $M = |-i\nabla|$. 
Then the operator $(|x|+i)^{-1}(M + i)^{-1} \in \Ii_{d+1}$. 
\end{lemma}
\pf Since the function $f(x)=(|x|+i)^{-1}$ is $L^{d+1}(\Rd)$ and the operator
in question is just $f(x)f(-i\nabla)$,  
the result follows
by an application of Theorem 4.1 in \cite{Sim3}, which gives an estimate 
$$
\|f(x)g(-i\nabla)\|_{\Ii_{p}} \leq 2\pi^{-\frac{d}{p}} \|f\|_p \|g\|_p.
$$
\qed

Let $V$ be an 
operator of multiplication by a function $V(x)$ on $L^2(\Rd)$ on its natural
domain and such that $V$ is bounded with respect to $H_0$ having relative bound smaller than $1$.  This means the operator $V(H +i)^{-1}$ is bounded.  
Then $H = H_0 + V$ is also self adjoint (Kato-Rellich theorem ) on the domain of 
$H_0$ and its spectrum is also bounded below. Writing 
$(H_0 +i)(H+i)^{-1} = I - V(H+i)^{-1}$, we see that 
$(H_0 +i)(H+i)^{-1}$ is also bounded.   
Let $P$ denote the operator of multiplication by the indicator 
function $\chi_\Lambda$ of a bounded region $\Lambda \subset \Rd$ on $L^2(\Rd)$.
Let $E_{H}( A )$ denote the spectral measure of a bounded borel set $A$, 
with respect to the (projection valued) spectral measure of $H$.  Then,

\begin{thm}\label{thmla1}
Consider $L^2(\Rd)$ and the operator $H_0 = -\Delta$. 
Let $V$ be an operator of multiplication by a function $V(x)$, such that
$V$ is relatively bounded w.r.t. $H_0$ with relative bound $c <1$  and consider
$H=H_0+V$.  Suppose either 
\begin{enumerate}
\item $d\leq 3$, then $PE_{H}(A)$ and $E_{H}P$ are Hilbert-Schmidt, so $PE_{H}(A)P$ is trace class
for any bounded borel set $A$.
\item Suppose $d\geq 1$ and suppose $V$ is bounded or 
$\frac{\partial}{\partial x_j}V, j=1,\dots,d$ are relatively bounded w.r.t $H_0$.
Then $PE_{H}(A)$ and $E_{H}(A)P$ are trace class.  
\end{enumerate}
\end{thm}
\pf (1) Writing $PE_{H}(A) = P (|x|+i)^d(|x|+i)^{-d} (H_0+i)^{-1}(H_0+i)(H+i)^{-1}(H+i)E_{h}(A)$,
we see that since all the factors are bounded, it is enough to show that
$(|x|+i)^{-d}(H_0+i)^{-1}$ is Hilbert-Schmidt.  The operator $(H_0+i)^{-1}$ is multiplication
by $(|\xi|^2+i)^{-1}$ after taking Fourier transforms and hence is in $L^2(\Rd), d\leq 3$. 
Therefore an application of Lemma \ref{leml1}, shows that the product is Hilbert-Schmidt.  

(2) We will prove that $PE_{H}(A) \in \Ii_1$, the proof for
$E_{H}(A)P$ is similar. By taking a compactly supported smooth function $\phi$ which is 
value $1$ on the closure of $A$, we have $\phi(H) E_{H}(A) = E_{H}(A)$. 
We will therefore show that $P\phi(H)$ is trace class for any compactly
supported smooth function $\phi$.
We also note that the function $H \phi(H)$ is again a function of the same type as $\phi$.

Further since $P$ is multiplication by compactly supported function  of $x$, $P(x^2+i)^{d}$
is bounded.  Therefore we will show that $(x^2+i)^{-d}\phi(H) \in {\mathscr I}_1$.

We prove this by induction.  Before we start, we note that if $M\in \Ii_p$ and $N$ is 
a bounded operator then $MN \in \Ii_p$. 

First consider $(x^2+i)^{-1}\phi(H)$.  We write this product as $(x^2+i)^{-1}(H+i)^{-1}(H+i)\phi(H)$
and consider (recalling $M = |-i\nabla|$),
\begin{equation}\label{eqnl7}
(x^2+i)^{-1}(H+i)^{-1} = (x^2+i)^{-1}(M+i)^{-1} (M+i)(H_0+i)^{-1}(H_0+i)(H+i)^{-1}.
\end{equation}
The product of the first two factors is in $\Ii_{d+1}$ (since $(|\xi|+i){-d-1}$ is integrable), 
by Lemma \ref{leml1}, the next two
factors form a bounded operator (which can be seen by taking Fourier transforms).
The final two factors form a bounded operator as argued before the lemma.  Therefore the entire product
is in $\Ii_{d+1}$. Since $(H+i)\phi(H)$ is bounded also, we get that $(x^2+i)^{-1}\phi(H) \in \Ii_{d+1}$. 

Now assume that $(x^2+i)^{-n}\phi(H) \in \Ii_{\frac{d+1}{n}}$, and show that  
$(x^2+i)^{-n-1}\phi(H) \in \Ii_{\frac{d+1}{n+1}}$. We write, $\psi(H) = (H+i)\phi(H)$, then
\begin{eqnarray}
& (x^2+i)^{-d-1} \phi(H) \\ & = (x^2+i)^{-d-1}(H+i)^{-1}(H+i)\phi(H) \nonumber \\ 
& (x^2+i)^{-1}[(x^2+i)^{-n}, (H+i)^{-1}]\psi(H) + (x^2+i)^{-1}(H+i)^{-1}(x^2+i)^{-n}\psi(H).\nonumber
\end{eqnarray}  
Using Theorem 2.8 (2.5b) in \cite{Sim3}, (which says $M \in \Ii_q, N\in \Ii_r \implies MN \in \Ii_p$ with
$\frac{1}{p} = \frac{1}{q}+\frac{1}{r}$), so using the induction hypothesis and the already proved
fact that $(x^2+i)^{-1}(H+i)^{-1} \in \Ii_{d+1}$, the last term is seen to be in $\Ii_{\frac{d+1}{n+1}}$.

So we concentrate on the first term.  
\begin{eqnarray}\label{eqnl8}
& (x^2+i)^{-1}[(x^2+i)^{-n}, (H+i)^{-1}]\psi(H) \\ &= (x^2+i)^{-1}(H+i)^{-1}[H,(x^2+i)^{-n}](H+i)^{-1}\psi(H) \nonumber \\
&= (x^2+i)^{-1}(H+i)^{-1}[H_0,(x^2+i)^{-n}](H+i)^{-1}\psi(H) \nonumber \\
&=  (x^2+i)^{-1}(H+i)^{-1} \times\nonumber \\ &\left(-4ni\sum_{j=1}^d P_j x_j(x^2+i)^{-1} -2d(x^2+i)^{-1} + 4d(n+1) x^2(x^2+i)^{-2}   \right)\nonumber \\ & \times (x^2+i)^{-n}\psi_1(H) \nonumber \\     
\end{eqnarray}
where we set $(H+i)^{-1}\psi(H) = \psi_1(H)$, where $P_j = -i\nabla_j$.  
\begin{eqnarray*}
& (x^2+i)^{-1}(H+i)^{-1}P_j \\ &= (x^2+i)^{-1}(H_0+1)^{-1/2}(H_0+1)^\half(H+i)^{-1}(H_0+1)^{\half}(H_0+1)^{-\half}P_j,
\end{eqnarray*}
and using Lemma \ref{leml1}, Lemma \ref{leml2} below, we see that this expression is in $\Ii_{d+1}$.  Induction
hypothesis gives $(x^2+i)^{-n}\psi_1(H) \in \Ii_{\frac{d+1}{n}}$.
Therefore combining these two facts we see that
$(x^2+i)^{-n-1} \phi(H) \in \Ii_{\frac{d+1}{n+1}}$.  
\qed

\begin{lemma}\label{leml2}
Suppose either $V$ is bounded or $(\frac{\partial}{\partial x_j}V) (H+i)^{-1}, j=1,\dots, d$ are 
bounded.  Then $(H_0+1)^\half(H+i)^{-1}P_j$ is a bounded operator for each $j=1,\dots d$.
\end{lemma}
\pf Consider the case when $\frac{\partial}{\partial x_j}V (H+i)^{-1}$ is bounded for each $j$.  Then
writing the expression using commutators
\begin{eqnarray*}
& (H_0+1)^{\half}(H+i)^{-1}P_j= (H_0+1)^\half P_j(H+i)^{-1} \\ &+ (H_0+1)^\half(H+i)^{-1}[P_j,H](H+i)^{-1} \\ &
= (H_0+1)^\half P_j(H+i)^{-1} + (H_0+1)^\half(H+i)^{-1} (\frac{\partial}{\partial x_j}V)(H+i)^{-1}.
\end{eqnarray*}
The boundedness of the first term was seen before since $(H_0+1)^\half P_j (H_0+1)^{-1}$ 
and $(H_0+1)(H+1)^{-1}$ are bounded. 
second term is bounded by the assumption on $V$ and the boundedness of $(H_0+1)^{-\half}$. 

Now consider the case when $V$ is bounded, then taking $f,g$ in the domain of $H_0$, we have
$$
\langle f, (H_0+1) f\rangle = \langle g, (H_0 +V +c+ 1) f \rangle - \langle g, (V+c) f \rangle,
$$
where $c$ is a positive constant such that $H + c+1$ is a positive operator (which is possible
since $H$ is bounded below).  Since $H + c+1$ is positive it has a unique square root, so using the
boundedness of $V$ and the above inequality, we obtain, for some finite $C$, 
$$
\|(H_0+1)^\half f\|^2 \leq \|(H+c)^{\half}f\|^2 + C \|f\| \leq D \|(H+c)^{\half}f\|^2.
$$
Taking $f=(H_0+1)^\half g, \|g\|=1$, for a set of $g$ coming from $C_0^\infty(\Rd)$, we see that 
$$
K \leq \|(H+c)^\half (H_0+1)^{-\half}g\|^2, ~ K>0, 
$$
$K$ independent of $g$. This shows that $(H+c)^\half(H_0+1)^{-\half}$ has a bounded inverse and
that its inverse $(H_0+1)^{-\half}(H+c)^\half$ and $(H+c)^\half(H_0+1)^{-\half}$
are both bounded (since $M$ bounded implies $M^*$ is also bounded).  
Therefore writing
\begin{eqnarray*}
& (H_0+1)^\half(H+i)^{-1}P_j \\ &= (H_0+1)^\half(H+c)^{-\half}(H+c)(H+i)^{-1}(H+c)^{-\half}(H_0+1)^\half
(H_0+1)^{-\half}P_j,
\end{eqnarray*}
we see that the left hand side is bounded. \qed

We are now ready to present examples where the theorems of the previous
section are applicable.  We fist give a few examples of models on the lattice.

Consider $L^2(\RR^d)$, $H_0 = \Delta$, $q(n), n\in \ZZ^d$,
i.i.d random variables with distribution $\mu$ having compact support. Let
$\Lambda$ denote the unit cube centred at $0 \in \Rd$ and $\Lambda(n)$
denote the unit cube centred at the point $n \in \ZZ^d$.  
Let $V_\omega
= \sum_{n \in \ZZ^d} q^\omega(n) \chi_{\Lambda(n)}$, where $\chi_A$
is the operator of multiplication by the indicator function of $A$. Then
taking 
$$
H_\omega^\pm = \Delta \pm V_\omega, 
$$
we see that, since $V_\omega$ is bounded for each $\omega$, the conditions
of Theorem \ref{thmla1} are satisfied. Further taking
$G = \Rd$ and $L = \ZZ^d$, $(U_x f)(y) = f(y-x), on ~ L^2(\Rd)$, $q_{_{T^m\omega}}(n) = q_{\omega}(n+m)$,   
in Hypothesis \ref{hyp1}, the spectral projections
$E_{H_\omega^\pm}( (a, b) )$  are covariant families in the sense of Definition \ref{hyp1}.  The Theorem \ref{thmla1} shows that $\chi_{\Lambda(0)} E_{H_\omega^\pm}( (a, b) ) \chi_{\Lambda(0)}$ is trace class whenever $(a, b)$ is a bounded interval.  
Hence we can define the density of states and the ILAC as in equations (\ref{eqnl00}) and (\ref{eqnl3}), by taking $P$ to be multiplication by $\chi_{_{\Lambda(0)}}$.  Therefore the Theorems \ref{thml1}, \ref{thml2} are valid in this
case.  

Our theorem covers models where the random potential has the following forms.
\begin{itemize}
\item $V^\omega(x) = \sum_{n \in \Z^d} q^\omega(n) u(x-n)$, $\{q(n)\}$ i.i.d.random variables
whose distribution has compact support and $u$ a nice function with $u(x-n)$ summable. 
\item An addition of a periodic background potential $W$ to the random potential above.   
\item Addition of magnetic fields.
\end{itemize} 
If in all these cases the density of states have Lifshitz tails behaviour at the band edges the same
is acquired by the ILAC at an appropriate energy level.

\begin{remark}
Let us remark that in the above examples we can even replace the Laplacian $-\Delta$ with
a real polynomial function $Q$ of $-i\nabla$ and the results go through, if for some
$R>0$, the polynomial satisfies 
$$ 
c_1 \|\xi\|^{2n} \leq Q(\xi) \leq c_2 \|\xi\|^{2n}, ~ |\xi| > R, c_1, c_2 >0. 
$$
\end{remark}

\endthebibliography
\end{document}